# Synthesis of nanoparticles in carbon arc: measurements and modeling

**Shurik Yatom, Alexander Khrabry, James Mitrani[1], Andrei Khodak, Igor Kaganovich, Vladislav Vekselman, Brent Stratton, Yevgeny Raitses.**

Princeton Plasma Physics Laboratory, Princeton University, NJ 08540, USA.

Address all correspondence to Shurik Yatom at **syatom@pppl.gov**

**Abstract**

This work studies the region of nanoparticle growth in atmospheric pressure carbon arc. Detection of the nanoparticles is realized via the planar laser induced incandescence (PLII) approach. Measurements revealed large clouds of nanoparticles in the arc periphery, bordering the region with high density of diatomic carbon molecules. Two-dimensional computational fluid dynamic simulations of the arc combined with thermodynamic modeling explain these results due to interplay of the condensation of carbon molecular species and the convection flow pattern. The results have shown that the nanoparticles are formed in the colder, outside regions of the arc and described the parameters necessary for coagulation.

**Introduction**

In the past decades, inorganic nanomaterials have gained a significant following in many scientific and engineering communities. The unique optical, electronic and mechanical properties of these materials are very attractive for a variety of potential applications [1-5]; therefore efficient pathways for production and modification of the nanomaterials are highly sought for. Synthesis of nanomaterials, referred below to as nanosynthesis, facilitated by plasma sources has become a staple technique for nanomaterial production [6-8]. Some nanomaterials can only be synthesized with plasma and for some, plasma approach is favored due to industrial scale yield, better selectivity and improved material characteristics [9]. Carbon arc is a widely used plasma source for making a variety of carbon nanomaterials of various structures such as fullerenes, carbon nanotubes, nano-horns, nanofibers, graphene, etc [10-19]. Particularly in the arc the single-walled nanotubes, nanoparticles and nano-horns are synthesized, in flight, in the gas phase, as opposed to the surface growth in chemical vapor deposition (CVD). The gas phase synthesis also happens in laser ablation experiments, however the arc is less expensive method and also has a significantly higher yield. Unfortunately, the growth mechanism of these nanostructures in gas phase is still poorly understood on both microscopic and atomistic levels, citing the lack of ability to monitor the steps in the synthesis processes, including nucleation and growth. Most of the understanding in this area comes from a post-growth, ex-situ evaluation of the nanostructures, nanoparticles and the attached impurities, using various laser and X-ray spectroscopic techniques and a high-resolution electron microscope [20-22], together with a trial and error process on the experimental side: varying the catalyst [23], feedstock [24], background gas composition [25-26] etc [27].

---

[1] Currently in Lawrence Livermore National Laboratory, Livermore, CA 94550, USA.

In-situ diagnostics of the plasma-synthesis processes are needed to detect and characterize the structures as they are formed. Recent efforts have shown a successful utilization of pulsed lasers, producing laser-induced incandescence (LII) [28] or scattering photons from particles trapped in laser interference pattern, (coherent Raleigh Brillouin scattering approach [29]) to measure sizes of particles in a high-pressure carbon arc environment. Combined with the traditionally established methods for diagnostics of plasmas and its species [30-31], we can now establish an improved picture of how and where the nano-structures are assembled and what atomic and molecular species are important for the synthesis purposes.

The LII diagnostics has shown a very promising potential for nanoparticle characterization, as it relies on relatively common equipment and is easy to implement [32]. In the combustion and in aerosol community LII is employed for a few decades now, for measurements of soot in flames or polluted atmosphere. In low to mid-pressure plasma LII was also utilized to a certain success, expanding the approach to consider scattering particles, larger than the nano-sized soot in flames and engines [33-36]. Previously, we have demonstrated how to adopt the LII model to high-pressure plasma environment [37-38] and deduce the particle sizes from the temporal evolution of the LII pattern [28]. In this work we present the results of imaging the incandescence, which allows us to detect large concentration of nanoparticles in the regions outside the arc. The experimental findings are validated with the results of fluid dynamics simulations, which explain these findings due to the formation of nanoparticles by condensation of carbon molecular species in these regions.

**Experimental setup**

The carbon arc setup used in this work has been previously described in detail in Ref. 28. Two graphite electrodes are aligned vertically, with the cathode on the top and the anode at the bottom. The diameters of the electrodes are ~1.1 cm and 0.6 cm for the cathode and the anode, respectively. The experimental chamber is evacuated with a mechanical pump and subsequently filled with He gas to a pressure of about 500 Torr. The discharge was maintained with a constant current of 60 A. To ignite the discharge the anode is biased with 100 V and gradually brought in contact with the cathode via mean of mechanical stepper motor. Once the electrodes are electrically "shortened", the stepper motor gradually lowers the anode position, adjusting the inter-electrode gap in order to maintain the voltage in the set range of 25-30 V. The LII diagnostics is realized via the use of the Quantel Nd:YAG laser, shooting a 1064 nm laser beam with full width at half maximum (FWHM) duration of 8 ns, with a maximum energy of 100 mJ, at frequency of 0.5 Hz. The laser beam is shaped with a cylindrical lens, to produce a laser sheath with vertical length of ~0.35 cm and width of ~0.05 cm. The beam path is 3 mm off axis of the anode, to the direction of the chamber view port; so that the light from the particles would not be obscured by the electrodes (see Fig.1.). The incandescence is recorded with an iCCD camera (Andor iStar), equipped with an optical filter (central wavelength = 633 nm, FWHM=10 nm). The central wavelength is selected in order to avoid the molecular and atomic emission lines of carbon and helium. The camera operation is synchronized with the laser pulse, by means of the digital delay generator (BNC 575) and the exposure of camera is set to 200 ns, in order to capture the peak of the incandescence signal, occurring after the rapid heating by the laser. Additionally, 200 ns is short enough time, so that the constant radiation from the plasma and the electrodes does not saturate the sensor. Let us further refer to this approach towards LII as planar LII (PLII). We have also employed a fast-framing camera

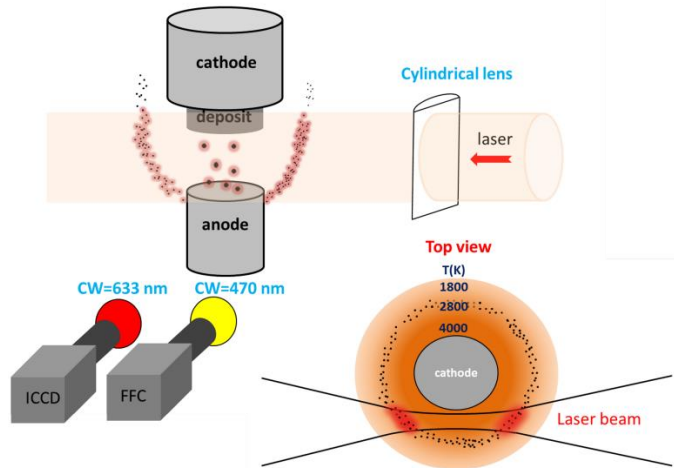

*Figure 1. Experimental arrangement for PLII detection, spectral imaging and the laser beam intersection with the region of interest.*

(Phantom 7) in order to monitor the behavior of the arc [39]. In this case we use an optical filter with central wavelength of 470 nm, where the Swan band emission ($C_2$) is located. Fast framing camera was triggered to acquire 50 frames during 1 ms, for each laser shot and the corresponding ICCD exposure

(PLII image). The frame exposure time was 1 µs and the delay between frames was 20 µs, when 40 first frames were acquired before the laser and 10 after it. Essentially the ICCD camera is recording the LII image and the fast-framing camera is monitoring the arc behavior leading up to and following the laser shot. Both cameras are observing the discharge from the same viewing port; however the images are obtained at a slightly different angle, thus exhibiting a minor deviation in the image, despite being captured simultaneously.

**Arc model**

The measurements presented in the paper are complemented by the results of 2D-axisymmetric steady state simulations of the carbon arc discharge in helium atmosphere performed using computational fluid dynamics (CFD) code ANSYS CFX which was highly customized for this purpose. The arc model comprised the gas phase model coupled to the models of heat transfer and electric current in the electrodes [40]. The gas flow and carbon transport were modeled in the whole chamber, including the arcing volume and the chamber bulk. Free convection in the chamber due to gas heating by the electrodes was modeled as well as forced convection in the arcing volume due to evaporation of the anode material. Non-uniform ablation of the graphite from the anode surface and carbon deposition at the cathode were self-consistently determined from the energy balance at the surfaces of the electrodes. Multiple surface physics phenomena such as space-charge sheaths, recombination of plasma ions, electron work function and radiation, were taken into account in the surface energy balance for accurate prediction of the ablation and deposition areas and rates. Non-equilibrium plasma model was utilized: separate equations for temperatures of electrons and heavy particles were solved, diffusion of electrons and ions were accounted for, allowing better prediction of the electric field strength, heating and temperature profiles in the arcing volume. Transport coefficients in the plasma were taken from [41]. The plasma model was benchmarked in Ref. [42] by comparison to the results of previous numerical studies [41] and in Ref. 43 by comparison to analytical solutions. Convective-diffusive transport of atomic carbon in the helium atmosphere was solved yielding a profile of carbon atoms density in the arcing volume and surrounding areas. Local chemical composition of carbon–helium mixture was assumed to be equilibrium and was obtained using Gibbs free energy minimization approach similar to one utilized in Refs. [44], [31]. Partial pressure of carbon gas was limited by the temperature-dependent saturation pressure of carbon vapor taken from Ref. [45]. In other words, condensation equilibrium is assumed: once temperature decreases below saturation point, condensation of the carbon gas takes place equalizing the gas partial pressure with the saturation value. The results for plasma composition were validated by comparison of $C_2$ molecules density profile with the results of spectroscopic measurements using Laser-Induced Fluorescence (LIF) technique [31]; good agreement was obtained.

**Experimental results**

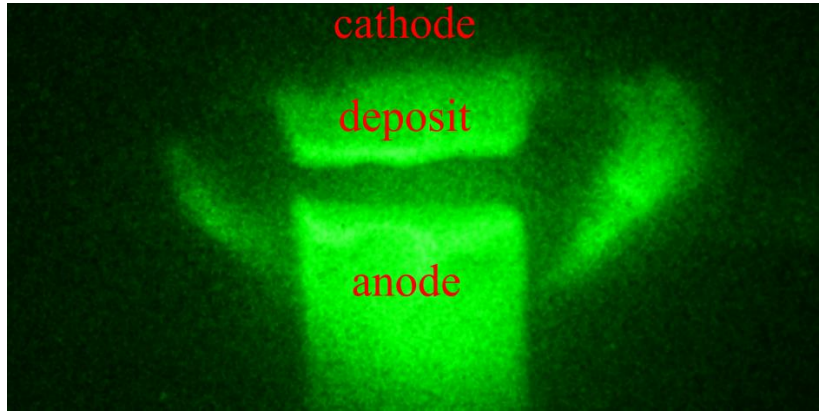

*Figure 2. A typical PLII image during the carbon arc discharge at the 500 Torr pressure helium gas. The incandescence from the regions of particles coagulation is identified on the right and the left side of the inter-electrode gap, assuming crescent-like formations. Additionally, the hot part of the anode and the hot deposit on the cathode are seen as well.*

Figure 2 shows an exemplary PLII image. The strongest feature of nanoparticle incandescence pattern appears on both sides of the electrode gap, in a form of partial, non-symmetrical crescents. Between the electrodes the signal can both be incandescence of the large graphite chunks or the radiation from the arc. This picture exemplifies the idea of two very distinct populations of particles that exist in different regions, echoing the results reported in Ref. [28]. Figure 3 shows the same PLII image superimposed with a series of the spectrally-filtered fast framing camera images. The fast-framing image presents the typical distribution of $C_2$ molecules around the arc core, resulting in a bubble-like structure. This "$C_2$ bubble" has been described in details earlier [30, 31], showing a larger carbon dimer density on the edges and a corresponding deficiency in the plasma channel. In Figure 3 the images depict the motion of this bubble-like structure from the left side of the anode to the right side and back, occurring in roughly 200 μs. The frame series is captured before, during and after the onset of the laser pulse and the resulting PLII image. The time marks on the frames are the times of the $C_2$ image capture and are relative to the laser pulse onset (and the subsequent PLII image acquisition) at t=0 μs.

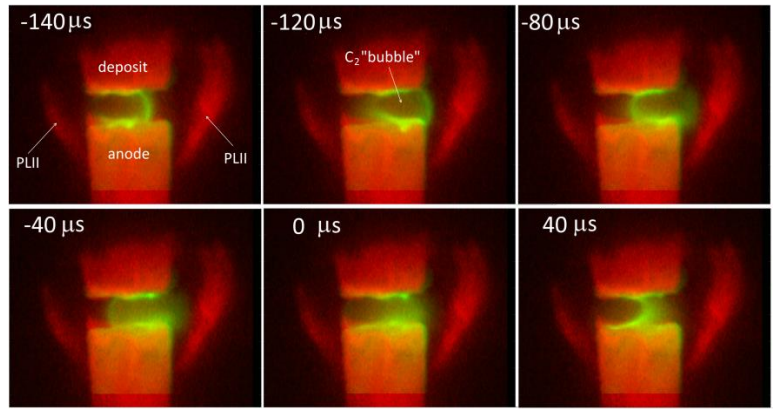

*Figure 3. PLII vs C₂. This collage illustrates the oscillation of the C$_2$ bubble-like structure, prior and after the acquisition of the PLII image. The PLII from ICCD camera image (red) and fast-framing camera image (yellow) are fused together. The times on the images in the collage are the times of fast-framing image (i.e. C$_2$ image acquisition), relatively to the laser shot time at t=0 µs, whereas the PLII image is the same on all frames and is obtained at t=0 µs.*

**Modeling results and their comparison with experiments**

The arc modeled was similar to one used in the experiments: 6 mm diameter cylindrical anode, 7 mm diameter cathode corresponding to the width of the carbonaceous deposit grown on the actual cathode, 2-3 mm inter-electrode gap, 60 A arc current. Results of the simulations are shown in Fig. 1. Density profiles of atomic and molecular carbon species are shown in Fig. 1a with different colors for different species. The color plots are cut-off for density values below certain levels to avoid overlapping of the plots and emphasize onion-like structure of the regions occupied by different species. In the arc center, where the gas temperature is at its highest and reaches 6500 K, see Fig. 1c, the carbon gas is almost completely dissociated and present in atomic form. Near the electrodes and at the arc periphery, about 3 mm from the axis of symmetry, the gas temperature decreases to about 4500 K, see Fig. 1c. In this bubble-shaped region, density of the diatomic carbon molecules is at its highest. Note that presence of this C$_2$ bubble is confirmed by the present experiments and by the experiments of Ref. [31]. Further away from the arc axis, at a distance of about 4 mm, the gas temperature is about 3500 K. In this region, the carbon gas is present mostly in a form of C$_3$ molecules.

Source of the carbon species in the arc is the ablation of the anode material. Flow pattern in the arc (Fig. 1b), indicates that most of the material ablated from the anode surface goes directly to the cathode front surface where it deposits. This picture is in accordance with the experimental observations of the deposit growth at the cathode and with the measurements of the ablation and deposition rates performed in Ref. [30,46].

However, some flow lines initiated from the anode front surface, closer to its side walls, deviate aside from the cathode, leave the inter-electrode gap and merge with relatively slow upwardly directed convective flow of helium along the side walls of the electrodes (Fig. 4b). These outgoing streamlines are also plotted in Fig. 4a over the density profiles. This flow carries some small fraction of the ablated material away from the arcing volume where it merges with the convective flow of helium and cools down. Eventually the gas pressure decreases below saturation point, and the carbon vapor starts to condense forming nanoparticles. As a result, the carbon gas density rapidly decreases along the streamlines, following the decrease of the saturation pressure which is a strong function of temperature (one order of magnitude decrease with temperature decrease by about 200 K) [45]. Let us define carbon gas supersaturation degree $S = p_C/p_{satur}$, with $p_C$ is the partial pressure of carbon in the carbon-helium mixture and $p_{satur}$ is the saturation pressure of carbon vapor. The condensation occurs for $S \geq 1$. Our estimates below show that the condensation is fast, keeping S close to unity. We can estimate the amount of the condensed material by calculating values of S, while the condensation is artificially kept from occurring. In Fig. 4 a, red lines show the locations of (i) S=1 (condensation starts), and (ii) S=10 (90% of the material are condensed). Note that according to our simulations, the formation of nanoparticles starts in the region where $C_3$ molecules are the dominant carbon species (Fig 4a).

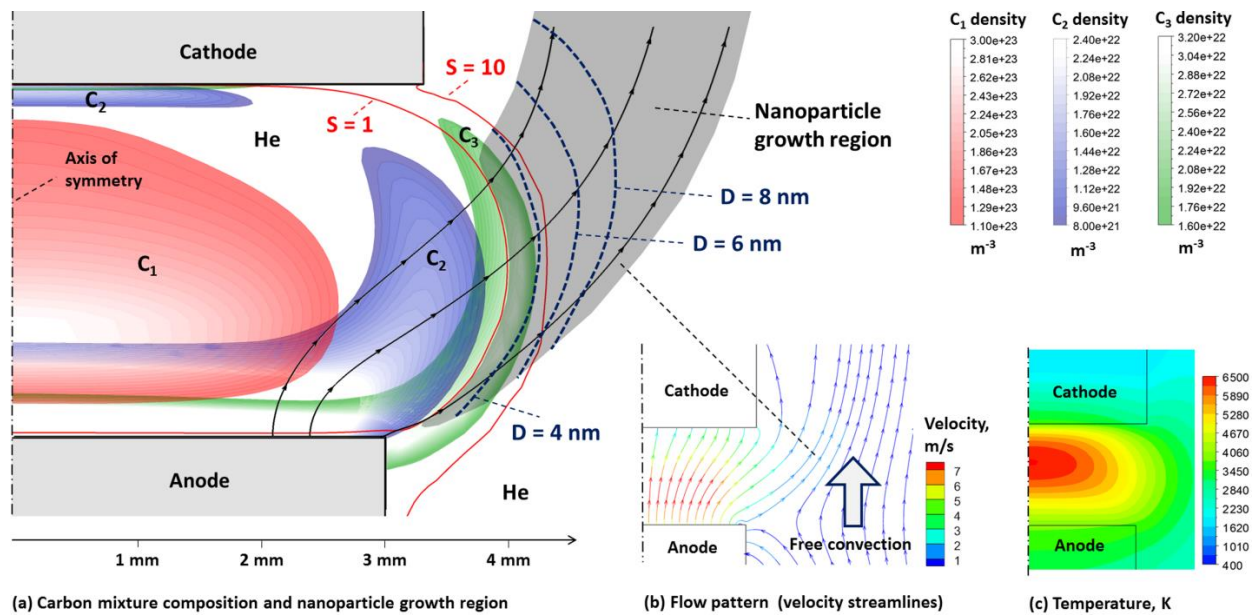

(a) Carbon mixture composition and nanoparticle growth region  
(b) Flow pattern (velocity streamlines)  
(c) Temperature, K

*Figure 4. Results of the carbon arc simulations. (a) Density profiles of various carbon species and nanoparticles growth region along selected flow streamlines; (b) flow pattern; (c) gas temperature profile. Red lines represent locations where carbon vapor supersaturation degree S equals to unity and to ten. At S=1, carbon vapor condensation and nanoparticle formation starts. By S=10 most of the vapor is condensed. Blue dashed lines show locations by which average diameter of the nanoparticles reaches 4, 6 and 8 nm due to their agglomeration.*

Growth of the nanoparticles can be described using a simple model of agglomeration of carbon clusters, similar to one used in Ref. [47]. Noteworthy, that in Ref. [47] $C_3$ molecules were also considered as a main precursor for growth of the nanoparticles. When growing, the nanoparticles are transported with the flow. Region of the nanoparticles growth is shown in Fig. 4a by a shaded area. Assuming that at temperature below saturation point all collisions of carbon molecules and clusters lead to their agglomeration, size distribution of the agglomerated particles should not be broaden, and a simple relation for variation of the particles density along their trajectory as they are carried by the steady state flow can be written as:

$$\frac{dn}{dt} = v\frac{dn}{ds} = -D^2 n^2 \sqrt{\frac{4\pi kT}{m}}. \tag{1}$$

Here, $n$ is density of carbon particles, including molecules and multi-atomic nanoparticles, $v$ is flow velocity, $s$ is a distance passed by the agglomerating particles from the location where the agglomeration started (S=1), $D=2r_0(n_0/n)^{1/3}$ is their average diameter, $m = m_{C_3}(n_0/n)$ is average mass, $T$ is the gas temperature, $r_0 \approx 1.2 \cdot 10^{-10} m$ is distance between atoms in a cluster (bond length), $n_0$ is initial density of the particles, i.e. density of $C_3$ molecules at the location S=1, $m_{C_3}$ is mass of a $C_3$ molecule. According to the simulations, $n_0$ is about $3 \cdot 10^{16} cm^{-3}$, see Fig 4.a. Solution of Eq. (1), assuming constant gas temperature, is:

$$n = \left(n_0^{-5/6} + \frac{5}{6}A\Delta t\right) \approx (A\Delta t)^{-6/5}. \tag{2}$$

Here, $A = (2r_0)^2 n_0^{1/6} \sqrt{\frac{4\pi kT}{m_{C_3}}}$.

This yields for average diameter of the nanoparticles:

$$D = 2r_0 n_0^{1/3}(A\Delta t)^{2/5}. \tag{3}$$

Spatial profiles of the nanoparticles density and average size in the growth region can be obtained by integration of Eq. (1) along the flow streamlines and are plotted in Fig. 5. Note that the nanoparticles diffusion has rather small effect on their motion compared to the convection and does not significantly influence the particles density, diameters and shape of the growth region. However, minor fraction of the particles can spread out from the sides of the growth region. Ratio of convection and diffusion time scales is:

$$\frac{t_{conv}}{t_{diff}} = \frac{LD_p}{vW^2} \approx 0.003.$$

Here, $L \approx 1\ cm$ is characteristic length of the nanoparticles growth region, $W \approx 3\ mm$ is its width (see Fig. 4), $D_p$ is particles diffusion coefficient defined as [25]:

$$D_p \approx \frac{(kT)^{1.5}}{6\sqrt{m_{He}}\ p\ D^2} \approx 3 \cdot 10^{-6}\ m^2/s,$$

where $p = 2/3\ atm$ is the background pressure, $m_{He}$ is mass of a helium atom.

Spatial profile of the nanoparticles average size in the growth region is shown in Fig. 4a by dashed dark blue lines. As evident from the Figs. 4 and 5, nanoparticles grow very fast in the beginning when they are small and their density and thermal velocities are high (assuming T=const): it takes a small fraction of a millimeter for the nanoparticles to gain average diameter of 3 nm corresponding to about 10 000 atoms per particle. Almost no small molecules are left in the mixture by this location. This is consistent with the previous statement of fast condensation of the carbon vapor. After one millimeter of path, the nanoparticles grow relatively slow gaining 3 nm in diameter in about 1 mm of path and reaching the average diameter of 20 nm 7 mm from the start of nucleation. Their density significantly decreases to about $5 \cdot 10^{10}\ cm^{-3}$ after they traveled 1 cm, but it is still sufficient for application of the LII technique. Note that according to Eq. (3), nanoparticles average diameter is rather weakly dependent on number density $n_0$ of carbon molecules and on the gas temperature.

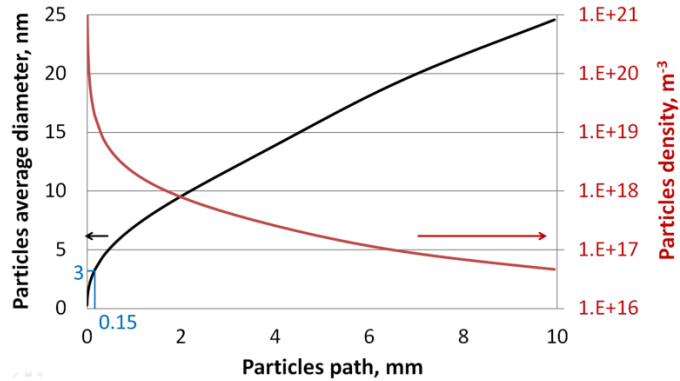

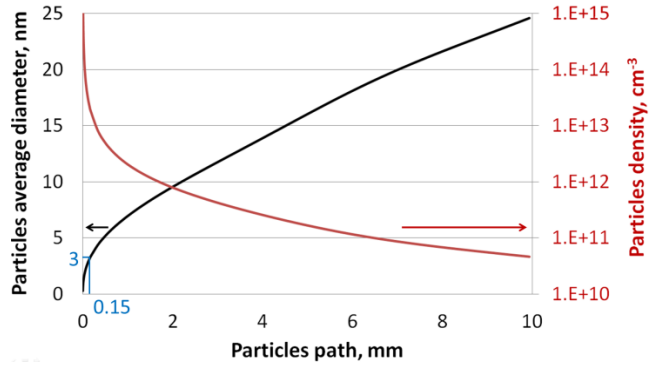

*Figure 5. Density (red line) and average diameter (black line) of the synthesized nanoparticles as a function of distance they passed from the beginning of carbon vapor condensation. At first 0.15 mm nanoparticles grow very fast and reach 3 nm corresponding to 10 000 atoms per particle. After that the growth rate decreases to about 3 nm in 1 mm of path.*

In Fig. 6, results of the simulations (white lines) are compared with the experimental photo of $C_2$ molecules emission and the nanoparticles incandescence. Regions occupied by $C_2$ molecules and by growing nanoparticles are shown by solid lines. Spatial distribution of the nanoparticles average diameter is shown by dashed lines. Note that in Ref. [31], the arc at lower currents was stable and symmetric allowing direct comparison between the measurements and results of 2D-axyssimmetric simulations for $C_2$ density profile. In the present experiments, the arc channel moves along the electrode surface at high frequency, accompanied by motion of the bubble-shaped region occupied by $C_2$ molecules, as shown in Fig. 3. This motion cannot be captured explicitly in the 2D-axisymmetric steady state simulations. In Fig. 6 the $C_2$ bubble is at its most right position having most effect on the nanoparticles growth region. Results of the simulations are shifted about 1 mm to the right to fit the experimental position of the $C_2$ bubble With this shift, location of the nanoparticle formation region also came into good agreement with the experimental measurements.

Simulated average diameter of the growing nanoparticles can be compared to experimental measurements [28] performed using TiRe LII technique. In the measurements, the signal was integrated along vertical line segments which are shown in the figure by yellow dashed lines. Results of the measurements are posted using yellow font. Good agreement of the particles sizes is observed: about 20 nm in the experiments versus about 15 nm in the simulations. Noteworthy that the TiRe LII measurements indicate 25 nm particles at distances of 8 mm and further from the axis of symmetry, where neither simulations nor the PLII photo show presence of the nanoparticles. This disagreement can be explained by the fact that the TiRe LII results are time averaged over the arc run. Oscillating behavior of the arc and dependence of the flow pattern on the inter-electrode gap size (which significantly changes during the arc run), can bring carbon further away from the electrodes. However, these effects are out of the scope of the paper, they will be considered in follow-up publications.

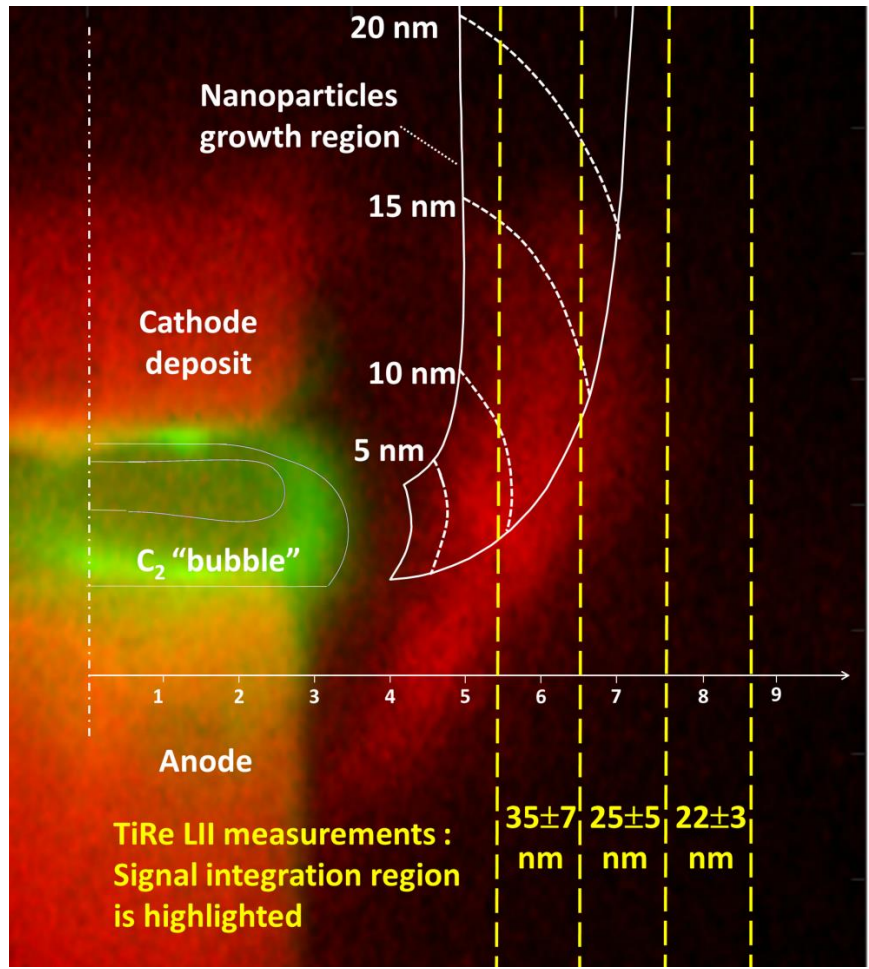

*Figure 6. The $C_2$ bubble and the nanoparticles growth region; results of the simulations are shown by white lines over the experimental photo. Locations where nanoparticles have certain average size are shown with white dashed lines. Yellow dashed lines indicate locations of the signal line integration in measurements of the nanoparticles average diameter [28] performed using TiRe LII technique. Areas from which a TiRe LII signal was collected are highlighted and the mean particle diameter for each area is shown.*

Let us examine the results demonstrated in Figure 6: the region of nanoparticle growth by coagulation roughly coincides with the area where PLII detects major concentration of nanoparticles. The calculated and the measured distributions of $C_2$ correspond very well too. The simulation does a great job showing the regions that allow the coagulation of nanoparticles, due to the existence of carbon feedstock and the sufficiently low temperatures. In the simulation all the particles are flushed upwards, with the gas flow. The experiment indeed shows the majority of the particles retracing this behavior, indicated by the well-defined "horns", however some diffusion in the radial direction occurs as well. This is expressed by the dampened incandescence beyond the "horn" structure (best seen at Figure 2) and also by the fact the TiRe LII was measured as far as 13 mm away from the arc axis [28].

A comparison between the nanoparticle sizes predicted by simple coagulation model and the ones measured in Ref. [28], by the means of TiRe LII can be summarized as follows. The TiRe LII signals were collected from segments 30 mm high in axial and 1.25 mm wide in radial directions. Therefore, the TiRe LII data offers a great spatial resolution in the radial direction, but not in axial direction. Additionally, the mean sizes of nanoparticles represent rather a broad distribution obtained for each radial segment. Finally, the TiRe LII results are also time-averaged over 150-200 signals measured during each arc run. In light of these remarks, the best correspondence between the coagulation and TiRe LII results are at 7 mm distance. At 6 mm the experimentally measured mean size is somewhat larger which can be explained by the following: the mean diameter in TiRe LII measurements is larger than in simulation, due to the occasional presence of large graphite chunks at the r=6 mm region. As we go further in radial distance the occurrence of these large chunks is less frequent, resulting in a slight decrease of the mean size of nanoparticles with distance. In the regions adjacent to the arc (r<6 mm) these chunks are frequent, hence the sizes measured there are much larger, completely overshadowing the contribution from the 5-10 nm particles, predicted in the calculation.

## Conclusions

This work shows that PLII method has successfully imaged the spatial distribution of nanoparticles, synthesized in the near-arc region, while mechanism of the nanoparticles formation and growth was revealed in the modeling performed in connection with 2D simulations of the arc. The majority of the observations from both the spectral imaging and the PLII were verified and reproduced in the modeling.

Modeling shows that formation of nanoparticles takes place in a narrow spatial region outside the arc, at a distance of ~5 mm from the hot arc core. The $C_3$ molecules serve as the building blocks for the nanoparticles. The agglomeration process happens faster than the variation of gas mixture composition, thus preventing formation of larger carbon molecules. Almost all carbon gas is condensed close to the region where nucleation begins, forming small nanoparticles with diameter ~3 nm. The later nanoparticle agglomeration is rather slow, resulting in nanoparticles with diameter of ~20 nm, at ~1 cm along the flow. Thermal convection of the ambient helium gas, heated by the contact with electrodes, plays crucial role in determining the shape of the nanoparticle growth region, while role of the particles diffusion is small.

These results will contribute to the development of accurate models for growth of nanoparticles and nanotubes in plasma environments.

A good agreement between the experimental data and the simulations on shapes of regions occupied by the nanoparticles and $C_2$ molecules as well as average size of the nanoparticles was obtained. These conclusions were obtained despite unstable behavior of the carbon arc due to rapid oscillations and the fact that the feedstock production via ablation is coupled with the plasma. For this reason, the PLII constitutes a seminal step in constructing a toolbox of experimental approaches that allows the in-situ exploration of nanoparticle formation.

### Acknowledgements

The authors would like to thank to Dr. Michael Schneider and Dr. Ken Hara for fruitful discussions and to A. Merzhevskiy for technical assistance

LII measurements and the thermodynamic simulations were supported by the U.S. Department of Energy (DOE), Office of Science, Basic Energy Sciences, Materials Sciences and Engineering Division The arc modeling was supported by the U.S. DOE Office of Science, Fusion Energy Sciences.